\documentclass{aer}

\usepackage[numbers,comma,round]{natbib}
\usepackage[figuresleft]{rotating}
\usepackage{txfonts}

\usepackage[utf8]{inputenc}
\usepackage[T1]{fontenc}
\usepackage{textcomp}
\usepackage{microtype}
\usepackage{enumitem}
\usepackage{setspace}
\usepackage[tableposition=top]{caption}
\usepackage[indentfirst=false,font+=small,leftmargin=0.4in,rightmargin=0pt,vskip=0pt]{quoting}
\usepackage{url}
\usepackage{hyperref}
\usepackage{graphicx}
\usepackage{placeins}

\usepackage{amsmath}
\usepackage{amssymb}
\usepackage[version=4]{mhchem}
\usepackage{siunitx}
\usepackage{longtable,tabularx}
\usepackage{graphics}
\usepackage{booktabs}
\usepackage{subcaption}
\usepackage{bm,xstring}
\usepackage{multirow}
\usepackage{pbox}
\usepackage{float}
\renewcommand{\copyright}{}

\begin{document}

\markboth{A. D. Menon et al.}{Three-Dimensional Impact Angle Control Guidance Law}

\title{Three-Dimensional Impact Angle Control Guidance Law with Field-of-View Constraint}

\author{Aswin D Menon\email{admenon@umich.edu}}
\affil{M.S.E Student, Department of Aerospace Engineering\\ University of Michigan\\ Ann Arbor, MI 48109 USA}

\author{Sayantan Pal\email{sayantanmech94@kgpian.iitkgp.ac.in}} 
\affil{Ph.D. Scholar, Unmanned Vehicle Systems Lab, Department of Aerospace Engineering\\ Indian Institute of Technology Kharagpur\\ Kharagpur 721302, India}

\author{Sikha Hota\email{sikhahota@aero.iitkgp.ac.in}}
\affil{Associate Professor, Department of Aerospace Engineering\\ Indian Institute of Technology Kharagpur\\ Kharagpur 721302, India}

\maketitle

\begin{abstract}
This paper presents a novel impact angle control guidance law to address the challenge of achieving any desired impact angle while satisfying the seeker's Field Of View (FOV) limit and maintaining acceleration continuity. The guidance problem is decomposed into two two-dimensional (2D) engagement planes. In the first plane, a pure proportional navigation guidance (PNG) law is employed toward a virtual target to align the missile velocity vector with the line-of-sight (LOS) to the target. In the second plane, a two-stage time-varying biased PNG (TVBPN) strategy is utilized to accurately shape the trajectory and achieve the desired terminal impact angle while maintaining compliance with FOV limitations. The position of the virtual target point ensures a smooth and continuous lateral acceleration at the transition between the first and second stages. The effectiveness and feasibility of the proposed law were validated through numerical simulations.
\end{abstract}

\section*{NOMENCLATURE}
\begin{deflist}
\listterm{$a$}{Speed of sound}
\listterm{$a_M$}{Lateral acceleration command}
\listterm{$b$}{Bias term}
\listterm{$B(t)$}{Bias integral}
\listterm{$B_{\text{ref}}$}{Reference value of the bias integral}
\listterm{$C_D$}{Total drag coefficient}
\listterm{$C_{D_0}$}{Zero-lift drag coefficient}
\listterm{$C_L$}{Lift coefficient}
\listterm{$D$}{Aerodynamic drag}
\listterm{$\mathbf{e}_{LOS}$}{Line-of-sight unit vector}
\listterm{$\mathbf{e}_m$}{Missile velocity unit vector}
\listterm{$\mathbf{e}_N$}{Unit normal to the engagement plane}
\listterm{$g$}{Gravitational acceleration}
\listterm{$h$}{Altitude}
\listterm{$k$}{Virtual target scaling factor}
\listterm{$K$}{Induced drag coefficient}
\listterm{$m$}{Mass}
\listterm{$M$}{Mach number}
\listterm{$N$}{Navigation gain}
\listterm{$Q$}{Dynamic pressure}
\listterm{$r$}{Relative range between missile and target}
\listterm{$R$}{Universal gas constant}
\listterm{$S$}{Reference area}
\listterm{$t$}{Time}
\listterm{$t_s$}{Switching instant}
\listterm{$T$}{Thrust}
\listterm{\textup{T}}{Atmospheric temperature}
\listterm{$V$}{Constant missile speed}
\listterm{$V_M$}{Time-varying missile velocity}
\end{deflist}

\subsection*{Greek Symbols}
\begin{deflist}
\listterm{$\gamma$}{Flight path angle / Heat capacity ratio}
\listterm{$\gamma_L$}{Critical impact angle}
\listterm{$\theta$}{Elevation angle}
\listterm{$\lambda$}{Line-of-sight (LOS) angle}
\listterm{$\rho$}{Atmospheric density}
\listterm{$\sigma$}{Seeker look angle}
\listterm{$\sigma_{\text{max}}$}{Maximum allowable look angle}
\listterm{$\tau_1, \tau_2$}{Time constants}
\listterm{$\psi$}{Azimuth angle}
\end{deflist}

\section{INTRODUCTION}
Effective missile guidance is pivotal for modern defence systems. Achieving a prescribed impact angle is crucial for maximizing warhead effectiveness and minimizing collateral damage. Since FOV (Field-of-View) constraints are inherent in any realistic seeker system, guidance laws are also required to ensure the target remains within the seeker's FOV throughout the engagement. Moreover, discontinuous acceleration commands can induce excessive structural loads on the missile airframe. Therefore, maintaining a continuous acceleration profile is essential for practical implementation.

In recent years, numerous studies have been conducted to design of impact angle control guidance (IACG) laws for planar engagements. Ratnoo and Ghose \cite{ratnoo2008impact} proposed a composite proportional navigation (PN)-based guidance law by combining an orientation phase with classical PN in the terminal phase to achieve any desired impact angle. This study was extended in \cite{ratnoo2010impact} to address the interception of nonstationary, nonmaneuvering targets. In \cite{ratnoo2016analysis}, FOV constraints were incorporated into a two-stage PN guidance (PNG) law while achieving all feasible impact angles. The same objectives were achieved with a reduced control effort using a three-staged PN guidance law by Ranjan et al. \cite{ranjan2019three}. Since these guidance laws are based on multiple PN stages, the resulting missile lateral acceleration profiles are discontinuous and exhibit abrupt jumps. Park et al. \cite{kim2013bias, park2017biased} designed a two-staged biased proportional navigation guidance (BPNG) law that considers FOV and lateral acceleration constraints. An energy-optimal impact angle control guidance law that explicitly incorporates FOV constraints for missiles equipped with strap-down seekers was presented in \cite{park2013optimal}. Lee et al. \cite{lee2013polynomial} proposed a time-to-go polynomial guidance (TPG) law while satisfying both terminal acceleration and seeker FOV constraints, where the acceleration command was formulated as a polynomial function of time-to-go. A time-varying two-staged BPNG law was designed by Yang et al. \cite{yang2016time} to achieve all possible impact angles while maintaining an FOV constraint. To further circumvent structural complexities, Zhang and Wang \cite{zhang2022new} proposed a guidance law featuring a time-varying navigation gain to achieve specific terminal angles without requiring phased switches or the addition of bias terms. Additionally, Gao et al. \cite{gao2017adaptive} investigated an adaptive terminal guidance strategy specifically focused on satisfying rigorous impact-angle constraints. However, all the aforementioned guidance laws are limited to two-dimensional engagement scenarios, which restricts their applicability to realistic three-dimensional (3D) interception problems.

To address the complexities of realistic missile-target interaction, recent research has extended impact angle control guidance into three-dimensional frameworks. Erer and Tekin \cite{erer2021impact} proposed guidance laws that extend biased PN to 3D engagement by rotating either the line of sight (LOS) or the velocity vector toward the desired terminal direction. In \cite{kim2021quaternion}, a quaternion-based guidance law was proposed that extends a 2D BPNG law to 3D by incorporating quaternion rotations between the missile velocity vector and the desired impact direction. Wang et al. \cite{wang2022three} presented a nonlinear guidance law that enables simultaneous control of impact angle and impact time using a two-stage structure, consisting of a sliding-mode-based IACG phase followed by a PNG phase. Another nonlinear guidance law, proposed in \cite{majumder2024three}, ensures interception within a predefined convergence time, using a sliding mode control framework and a predefined-time stability theory. Nanavati et al. \cite{nanavati2021lead} proposed a nonlinear guidance law that uses a finite-time state-dependent Riccati equation and a lead-angle formulation. Zhou et al. \cite{zhou2020three} developed an IACG law by decomposing the trajectory into two independently designed nominal curves, an elliptical path in the vertical plane and a Bézier curve in the lateral plane. Advanced nonlinear techniques, such as the fuzzy variable coefficients strategy combined with sliding mode control, have also been investigated by Xiong et al. \cite{xiong2016improved} to enhance the robustness of 3D guidance profiles. Moreover, recent advances have successfully incorporated learning-based adaptive manoeuver guidance \cite{wu2024reinforcement} and three-dimensional target manoeuver compensation algorithms \cite{qi2025integration} to tackle highly dynamic engagements. However, these 3D engagement studies do not explicitly consider the FOV constraints of the seeker.

In practical engagement scenarios, strict adherence to FOV constraints is indispensable due to the physical limitations of onboard seekers. Consequently, guidance laws must explicitly account for FOV limitations to ensure that the target remains within the seeker’s tracking capability throughout the engagement. Duvvuru et al. \cite{duvvuru2022three} proposed a guidance law that satisfies the terminal impact angle and the FOV constraints while maximizing the terminal speed and minimizing flight time, using an extended form of the generalized model predictive static programming technique. In \cite{gu2024three}, a guidance law was proposed that enforces prescribed performance and FOV constraints by shaping LOS angle errors through a range-dependent prescribed performance function. A predefined-time IACG law was introduced in \cite{wang2023three} that uses a novel predefined-time error dynamic tailored for second-order systems and auxiliary functions to prevent FOV violations. Wang et al. \cite{wang2022field} proposed an FOV-constrained guidance law that combines a PN term with a bias component shaped by a nonlinear error dynamic for speed-varying missiles. Another FOV-constrained guidance law \cite{han2022three} that constructs range profiles as cubic polynomials of LOS angles was proposed for speed-varying missiles. Addressing the challenge of diverse target kinematics, Du et al. \cite{du2026impact} recently introduced a sliding mode guidance scheme to enforce field-of-view limits against targets displaying various motion profiles. Hu et al. \cite{hu2020analytical} proposed a guidance law that achieves both impact angle and seeker FOV constraints using polynomial shaping of LOS angles in pitch and yaw, formulated as cubic polynomials of relative range, and a similar approach was reported in \cite{surve2024polynomial}. Ghosh et al. \cite{ghosh2017unmanned} proposed a composite PN-based guidance law ($\mu\alpha\text{CIG3D}$) that enables unmanned aerial vehicles to approach a stationary target from any arbitrary direction in 3D space while accounting for FOV limitations; however, this law results in multiple discontinuities in the lateral acceleration profile. Despite these developments, no existing 3D IACG framework simultaneously guarantees strict FOV constraint satisfaction, continuous acceleration profiles, and the ability to achieve arbitrary terminal impact directions.

Motivated by these gaps, this paper proposes a novel three-dimensional impact angle control guidance law for stationary targets that ensures strict FOV compliance while maintaining acceleration continuity. The key contributions are summarized as follows:
\begin{itemize}
    \item A novel 3D impact angle control guidance scheme is developed by decoupling the spatial engagement into two sequential planar phases, enabling target interception from any arbitrary spatial direction while strictly enforcing seeker FOV constraints.
    \item An analytical virtual target placement strategy is introduced for the primary alignment phase to completely eliminate the discontinuous lateral acceleration transients typically associated with multi-stage guidance transitions.
    \item A modified switching logic for the time-varying biased proportional navigation (TVBPN) law is formulated to ensure deterministic acceleration continuity and reliable phase initialization even for highly shallow terminal impact angles.
\end{itemize}

The remainder of this paper is organized as follows. The problem statement and a background review of planar guidance laws are established in Section~\ref{sec:prob_back}. Section~\ref{sec:law} details the formulation of the 3-staged 3D IACG scheme, including the modified TVBPN logic, virtual target placement, and analytical proof of FOV satisfaction. Numerical simulation results are presented in Section~\ref{sec:sims}, and Section~\ref{sec:concl} concludes the paper.

\section{PROBLEM STATEMENT AND BACKGROUND}
\label{sec:prob_back}

\subsection{Problem Statement}
\label{subsec:prob}

\begin{figure}[h]
\centering
\includegraphics[width=1\linewidth]{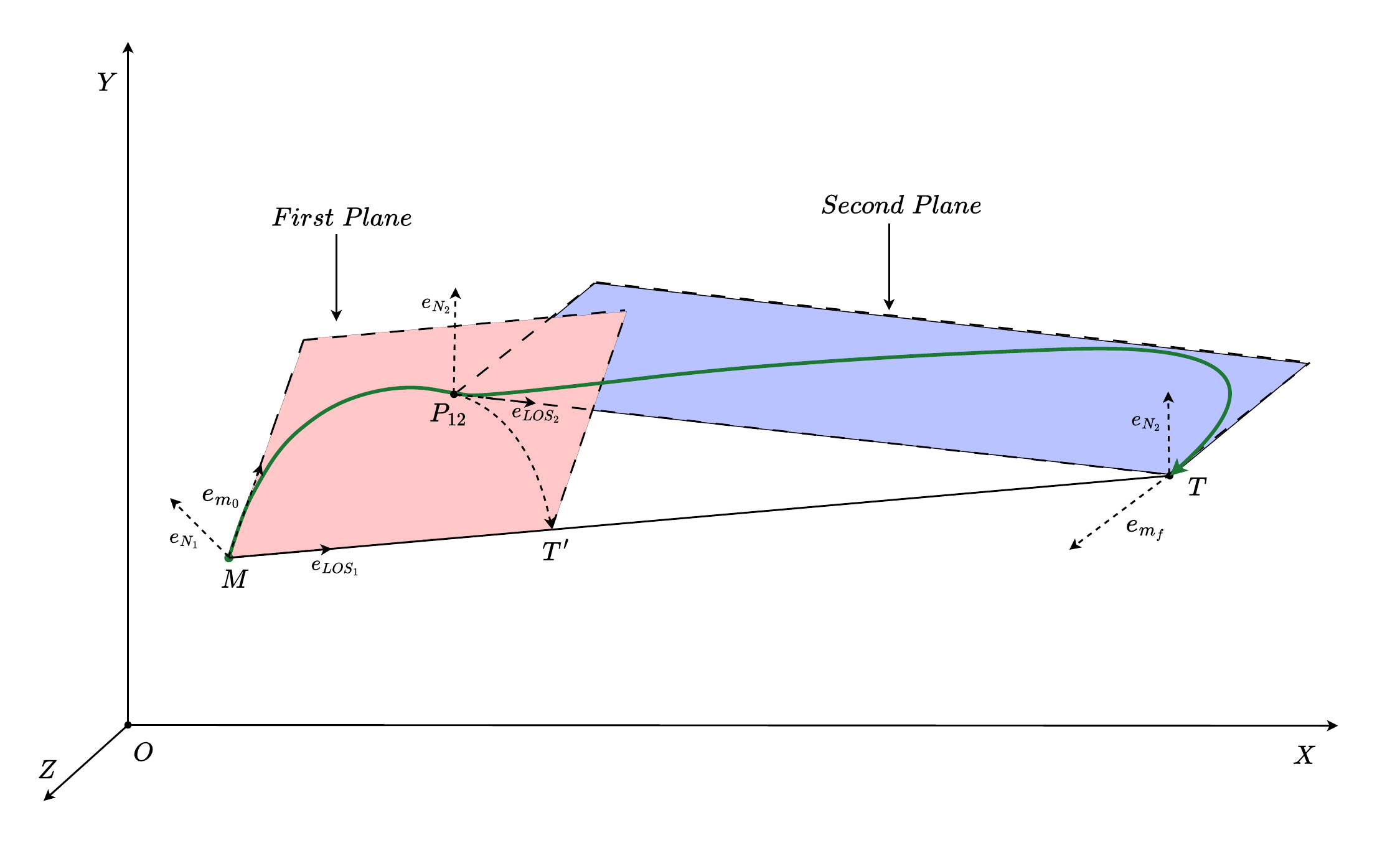}
\caption{\label{fig1} 3D Engagement Scenario}
\end{figure}

Consider an inertial coordinate frame $(X, Y, Z)$ centered at an origin $O$. Let $M$ and $T$ denote the initial spatial coordinates of the missile and a stationary target, respectively, as shown in Fig.~\ref{fig1}. The three-dimensional engagement is decoupled into two sequential planar phases. The first plane is defined by the initial line-of-sight (LOS) unit vector, $\mathbf{e}_{LOS_1}$, and the initial missile velocity unit vector, $\mathbf{e}_{m_0}$. The missile targets a virtual point $T'$ along the initial LOS until its velocity vector aligns with the LOS at the transition point $P_{12}$. The second plane is formed by the LOS unit vector at transition, $\mathbf{e}_{LOS_2}$, and the desired terminal impact unit vector, $\mathbf{e}_{m_f}$. The unit normals to the first and second engagement planes are respectively defined as:

\begin{equation}
\label{probeq1}
\begin{aligned}
\mathbf{e}_{N_1} &= \mathbf{e}_{LOS_1} \times \mathbf{e}_{m_0}, \\
\mathbf{e}_{N_2} &= \mathbf{e}_{m_f} \times \mathbf{e}_{LOS_2}
\end{aligned}
\end{equation}
Local 2D Cartesian frames are established for each plane to formulate the guidance problem. The first frame originates at $M$, with the principal axes aligned along $\mathbf{e}_{LOS_1}$ and $\mathbf{e}_{N_1}\times\mathbf{e}_{LOS_1}$. The second frame originates at the switching point $P_{12}$, with axes along $\mathbf{e}_{LOS_2}$ and $\mathbf{e}_{N_2}\times\mathbf{e}_{LOS_2}$.

\begin{figure}[h]
\centering
\includegraphics[width=0.7\linewidth]{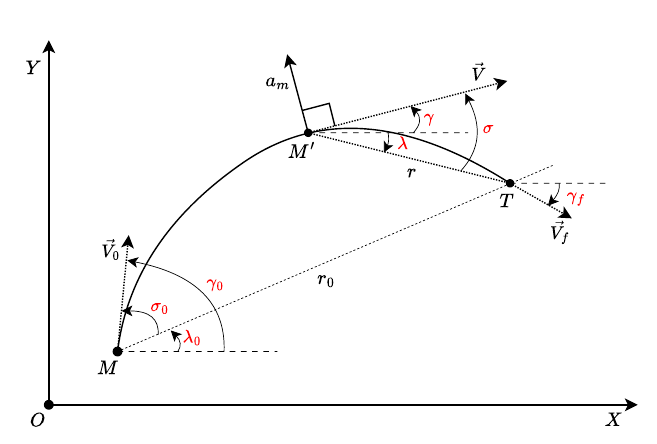}
\caption{\label{fig3} 2D Engagement Geometry}
\end{figure}

Since the three-dimensional engagement is decomposed into sequential planar sub-engagements, the governing dynamics within each phase can be rigorously described using standard planar interception kinematics, as shown in Fig.~\ref{fig3}. Let $V$ denote the constant missile speed, $r$ the relative range between the missile and the target, $\gamma$ the flight path angle, $\lambda$ the line-of-sight (LOS) angle, and $\sigma = \gamma - \lambda$ the seeker look angle. The corresponding planar kinematic relations are given by:
\begin{equation}
\label{2dkineqs}
\begin{aligned}
\dot{r} &= -V\cos\sigma, \\
\dot{\lambda} &= -\frac{V\sin\sigma}{r}, \\
\dot{\gamma} &= \frac{a_M}{V}
\end{aligned}
\end{equation}

\subsection{Background}
\label{subsec:back}

For a pure proportional navigation (PN) guidance law with navigation gain $N$, the lateral acceleration command $a_M$ is given by
\begin{equation}
\label{probeq11}
a_M = N V \dot{\lambda}
\end{equation}
A biased proportional navigation (BPN) guidance law incorporates an additive bias term $b$ to shape the trajectory, yielding the following lateral acceleration command:
\begin{equation}
\label{probeq12}
a_M = N V \dot{\lambda} + V b
\end{equation}
Using \eqref{2dkineqs}, the flight path angle rate is expressed as
\begin{equation}
\label{gamma_dot}
\dot{\gamma} = N \dot{\lambda} + b
\end{equation}
Integrating \eqref{gamma_dot} over the engagement interval $t \in [0, t_f]$ yields the terminal flight path angle:
\begin{equation}
\label{probeq14}
\gamma_f - \gamma_0 = N(\lambda_f - \lambda_0) + \int_{0}^{t_f} b\,dt
\end{equation}

For successful interception, the terminal condition $\gamma_f = \lambda_f$ must be satisfied. Enforcing this constraint yields the required reference value of the bias integral, $B_{\text{ref}}$:

\begin{equation}
\label{probeq15}
B_{ref} = \int_{0}^{t_f} b\,dt = (1 - N)\gamma_f + N\lambda_0 - \gamma_0
\end{equation}

\subsubsection{Review of the TVBPN Guidance Law}

To address 2D FOV-constrained IACG problems, the time-varying biased proportional navigation (TVBPN) law \cite{yang2016time} utilizes two distinct time-varying bias terms, $b_1$ and $b_2$, governed by a switching logic. During the initial phase, the bias $b_1$ drives the look angle $\sigma$ toward its allowable limit $\sigma_{\text{max}}$:
\begin{equation}
\label{law16}
b_1 = (1 - N)\dot{\lambda} + \frac{1}{\tau_1}(\sigma_{\text{max}} - \sigma)
\end{equation}
where, $\tau_1$ is the time constant. Substituting $b_1$ into the BPN acceleration command yields:
\begin{equation}
\label{law17}
a_m = V\dot{\lambda} + \frac{V}{\tau_1}(\sigma_{\text{max}} - \sigma)
\end{equation}
The subsequent phase utilizes $b_2$ to drive the bias integral $B(t)$ to the required reference value $B_{\text{ref}}$ necessary for the desired terminal impact angle:
\begin{equation}
\label{law18}
b_2 = \dot{B}(t) = \frac{B_{\text{ref}} - B(t)}{\tau_2}
\end{equation}
\begin{equation}
\label{law19}
\implies b_2 = \frac{(B_{\text{ref}} - B(t_s)) \exp\left(-\frac{t - t_s}{\tau_2}\right)}{\tau_2}
\end{equation}
where, $\tau_2$ is the convergence time constant and $t_s$ denotes the switching instant. The corresponding lateral acceleration command is given by
\begin{equation}
\label{law20}
a_M = N\dot{\lambda} + \frac{B_{\text{ref}} - B(t)}{\tau_2}
\end{equation}
The transition between the two bias formulations is governed by their relative magnitudes, with the switching logic defined as:
\begin{equation}
b = \begin{cases} b_1, & \left|b_1\right| < \left|b_2\right| \\ b_2, & \left|b_1\right| \geq \left|b_2\right| \end{cases}\label{law21}
\end{equation}

\section{3-STAGED 3D IACG LAW}
\label{sec:law}
The proposed 3D IACG scheme guarantees prescribed terminal impact angles, strict seeker FOV compliance, and continuous lateral acceleration. The three-dimensional engagement is decoupled into a pure PNG phase on the primary plane, followed by a two-stage time-varying BPN (TVBPN) phase on the secondary plane. To eliminate lateral acceleration discontinuities during the planar transition, a virtual target is strategically positioned on the initial LOS.

\subsection{Design of the 3-staged 3D IACG law}

The missile operates at a constant speed $V$, with control authority exerted exclusively via lateral acceleration. During the first stage on the primary plane $(X_1, Y_1)$, a pure PNG law with a navigation gain of $N=2$ is applied to drive the missile toward a virtual target located at $T'$ on the initial LOS, as shown in Fig. \ref{fig1}. The corresponding acceleration command and flight path angle rate are given by:

\begin{equation}
\label{law24}
\begin{aligned}
a_{M_1} &= 2 V \dot{\lambda}_1, \\
\dot{\gamma}_1 &= 2 \dot{\lambda}_1
\end{aligned}
\end{equation}

where, the subscript `$1$' denotes variables in the primary plane. For $N = 2$, the look angle rate reduces to $\dot{\sigma}_1 = \dot{\lambda}_1$. This phase continues until the missile velocity vector aligns with the line-of-sight (LOS) direction in the inertial frame (i.e., $\gamma = \lambda$). Based on the frame definitions, the initial conditions are given by:
\begin{equation}
\label{law27}
\begin{aligned}
\sigma_{1_0} &= \gamma_{1_0}, \\
\lambda_{1_0} &= 0
\end{aligned}
\end{equation}
Upon alignment of the velocity vector with the LOS, the engagement transitions to the secondary plane $(X_2, Y_2)$, where the initial conditions reset relative to the new LOS:

\begin{equation}
\label{law30}
\begin{aligned}
\sigma_{2_0} &= 0, \\
\lambda_{2_0} &= 0, \\
\gamma_{2_0} &= 0
\end{aligned}
\end{equation}
The lateral acceleration and the flight path angle rate in the secondary plane are governed by an augmented TVBPN law with navigation gain $N_2$:

\begin{equation}
\label{law28}
\begin{aligned}
a_{M_2} &= N_2 V \dot{\lambda}_2 + V b, \\
\dot{\gamma}_2 &= N_2 \dot{\lambda}_2 + b
\end{aligned}
\end{equation}
This yields a look angle rate of $\dot{\sigma}_2 = (N_2-1)\dot{\lambda}_2 + b$. Here, the subscript `$2$' denotes variables associated with the secondary plane.

\subsection{Modified TVBPN Switching Logic}

The standard TVBPN guidance logic assumes $\left|b_1\right| < \left|b_2\right|$ initially. However, for small desired terminal impact angles (typically $<5^\circ$), the required bias integral $B_{\text{ref}}$ becomes sufficiently small, causing $\left|b_2\right| < \left|b_1\right|$ at the onset of the secondary plane. If $b_2$ is selected initially, enforcing acceleration continuity during the planar transition becomes analytically intractable, since $b_2$ depends explicitly on the terminal condition.

To address this limitation, the switching logic is modified to ensure $b_1$ is strictly enforced as the initial bias of the secondary phase. Define a critical impact angle $\gamma_L$ such that the initial bias magnitudes are equal, i.e., $|b_{1_0}| = |b_{2_0}|$. Using the initial conditions in \eqref{law30}, along with \eqref{law16} and \eqref{law18}, this condition reduces to:

\begin{equation}
\label{law33}
\begin{aligned}
\left|(1 - N_2)\dot{\lambda}_2 + \frac{1}{\tau_1}(\sigma_{\text{max}} - \sigma_2)\right|
&= \left|\frac{B_{\text{ref}} - B(t)}{\tau_2}\right| \\[6pt]
\implies \quad
\frac{\sigma_{\text{max}}}{\tau_1}
&= \frac{(1 - N_2)\gamma_L}{\tau_2} \\[6pt]
\implies \quad
\gamma_L
&= \frac{\sigma_{\text{max}}}{(1 - N_2)} \frac{\tau_2}{\tau_1}
\end{aligned}
\end{equation}

The switching criterion is subsequently modified into a piecewise condition based on the desired terminal angle $\gamma_f$:

\begin{equation}
\label{law35}
b =
\begin{cases}
b_1, & |\gamma_f| < |\gamma_L|, \\[6pt]
\begin{cases}
b_1, & |b_1| < |b_2|, \\
b_2, & |b_1| \geq |b_2|
\end{cases}, & |\gamma_f| \geq |\gamma_L|.
\end{cases}
\end{equation}

This structured modification guarantees that $b_1$ initiates the secondary phase, thereby enabling deterministic enforcement of acceleration continuity while maintaining compliance with the FOV constraint.

\subsection{Virtual Target Placement}
The position of the virtual target is calculated  by enforcing continuity of the lateral acceleration magnitude across the planar transition, thereby avoiding structural loading transients:
\begin{equation}
\label{law36}
|a_{M_1}(t_{s_1})| = |a_{M_2}(0)|
\end{equation}
where, $t_{s_1}$ denotes the transition time from the primary to the secondary plane. Substituting the respective acceleration commands yields:
\begin{equation}
\label{law37}
|2 V \dot{\lambda_1}(t_{s_1})| = |N_2V\dot{\lambda_2}(0) + Vb_1(0)|
\end{equation}
For a pure PNG law with $N = 2$ against a stationary target, the missile follows a circular trajectory, resulting in constant lateral acceleration. Consequently, the acceleration at the transition instant equals its initial value in the primary phase. By defining the virtual target $T'$ at a fractional distance $k$ along the initial relative range $r_0$, the continuity condition simplifies to:  
\begin{equation}
\label{law38}
\frac{2 V^2 sin(\gamma_{1_0})}{k r_0} = \frac{V \sigma_{\text{max}}}{\tau_1}  
\end{equation}
Solving for the virtual target scaling factor $k$ yields:

\begin{equation}
\label{law39}
k = \frac{2 V \tau_1 sin(\gamma_{1_0})}{r_0 \sigma_{\text{max}}}
\end{equation}

\subsection{Satisfaction of the FOV limit}
To ensure continuous target tracking, the seeker look angle must remain strictly bounded throughout the engagement. In the primary phase, a pure PNG law with $N = 2$ is employed. As established in \cite{ratnoo2016analysis}, for $N > 1$, the look angle dynamics satisfy:

\begin{equation}
\label{law40}
\dot{\sigma}_1 = 
\begin{cases}
> 0 & \forall\ -\pi < \sigma_1 < 0,\ N > 1 \\
< 0 & \forall\ 0 < \sigma_1 < \pi,\ N > 1 \\
= 0 & \text{if } \sigma_1 = 0,\ \pi \\
= 0 & \text{if } N = 1
\end{cases}
\end{equation}
i.e., for any $N>1$ and positive initial look angle, the look angle rate is strictly negative ($\dot{\sigma}_1 < 0$), ensuring $\sigma_1$ monotonically approaches zero.

Upon transition to the secondary plane, the look angle dynamics are governed by the modified TVBPN law. During the $b_1$ phase:
\begin{equation}
\label{law41}
\dot{\sigma}_2(t) = (N_2 - 1) \dot{\lambda}_2 + b_1 = \frac{1}{\tau_1} (\sigma_{\text{max}} - \sigma_2)
\end{equation}
which implies exponential convergence of $\sigma_2$ toward the upper bound $\sigma_{\text{max}}$. Once the switching condition activates the $b_2$ phase at time $t_{s_2}$, the look angle rate becomes:

\begin{equation}
\label{law42}
\dot{\sigma}_2(t) = (N_2 - 1)\dot{\lambda}_2 + b_2
= \frac{(1 - N_2)V \sin \sigma_2(t)}{r(t)} + \frac{(B_{\text{ref}} - B(t_{s_2})) \exp\left( -(t - t_{s_2})/\tau_2 \right)}{\tau_2}
\end{equation}
As shown in \cite{yang2016time}, this condition ensures that $\dot{\sigma}_2(t) \leq 0$ for $t \geq t_{s_2}$, implying that $\sigma_2$ remains bounded and does not exceed $\sigma_{\text{max}}$. Therefore, the seeker FOV constraint is satisfied throughout all phases of the engagement.

\section{NUMERICAL ANALYSIS}
\label{sec:sims}

\begin{figure}[h]
\centering
\includegraphics[width=0.5\linewidth]{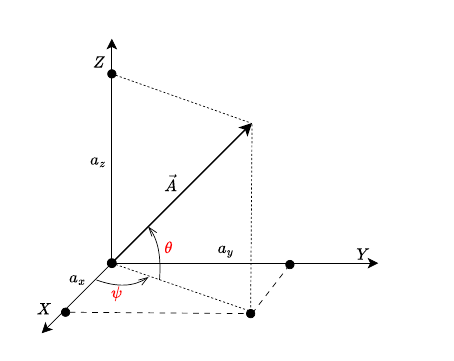}
\caption{\label{fig2} 3D Angles}
\end{figure}

The performance of the proposed 3D IACG scheme is evaluated through numerical simulations. Unless otherwise specified, the missile and target are initially located at $[0, 0, 0]^T$ m and $[2500, 2500, 2500]^T$ m, respectively. The missile maintains a constant speed of $50$ m/s. The maximum seeker FOV is constrained by a maximum allowable look angle of $\sigma_{\text{max}} = 45^\circ$. The guidance parameters are set to $N_2 = 3$, $\tau_1 = 2$ s, and $\tau_2 = 0.5$ s. The initial velocity vector is chosen such that the initial look angle satisfies $\sigma_0 < \sigma_{\text{max}}$. The spatial orientation of an arbitrary vector $\vec{A} = [a_x, a_y, a_z]^T$ is parameterized by its azimuth angle $\psi \in [-\pi, \pi]$ and elevation angle $\theta \in [-\pi/2, \pi/2]$, as illustrated in Fig.~\ref{fig2}, and defined as:
\begin{equation}
\label{probeq2}
\begin{aligned}
\psi &= \tan^{-1}\!\left(\frac{a_y}{a_x}\right), \\
\theta &= \tan^{-1}\!\left(\frac{a_z}{\sqrt{a_x^2 + a_y^2}}\right)
\end{aligned}
\end{equation}

\subsection{Results with Varying Impact Angles}

To demonstrate terminal angle control authority, the missile is launched with $(\psi_0, \theta_0)$: $(45^\circ, 0^\circ)$ and commanded to intercept the target at five distinct spatial orientations $(\psi_f, \theta_f)$: $(-45^\circ, 70^\circ)$, $(-60^\circ, -30^\circ)$, $(-30^\circ, -60^\circ)$, $(-90^\circ, 45^\circ)$, and $(-179^\circ, -75^\circ)$.

\begin{figure}[h!]
    \centering
    \subfloat[\centering Trajectory]{
    \includegraphics[width=0.6\linewidth]{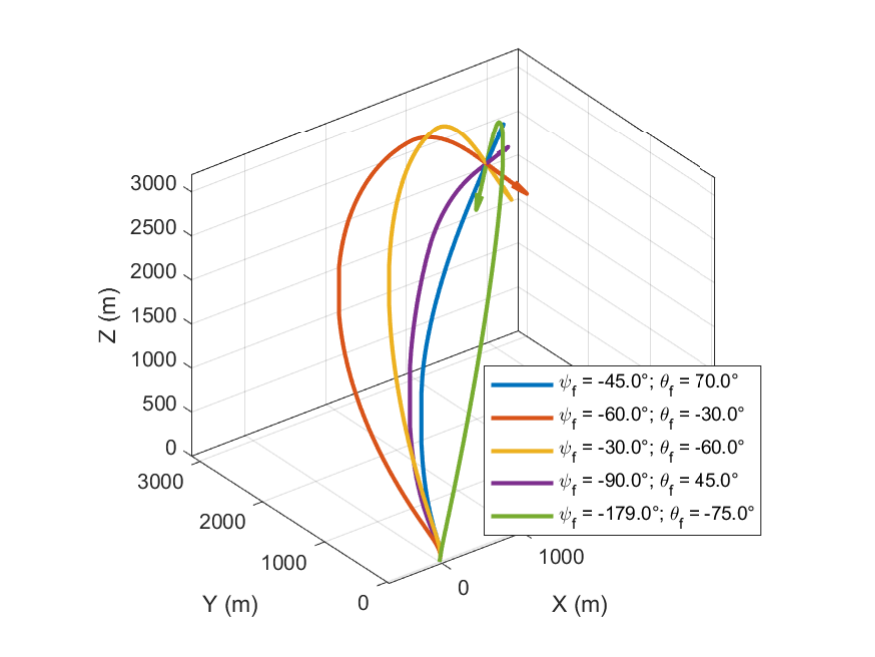}
    \label{s1:traj}
    }\\
    \subfloat[\centering Acceleration]{
    \includegraphics[width=0.5\linewidth]{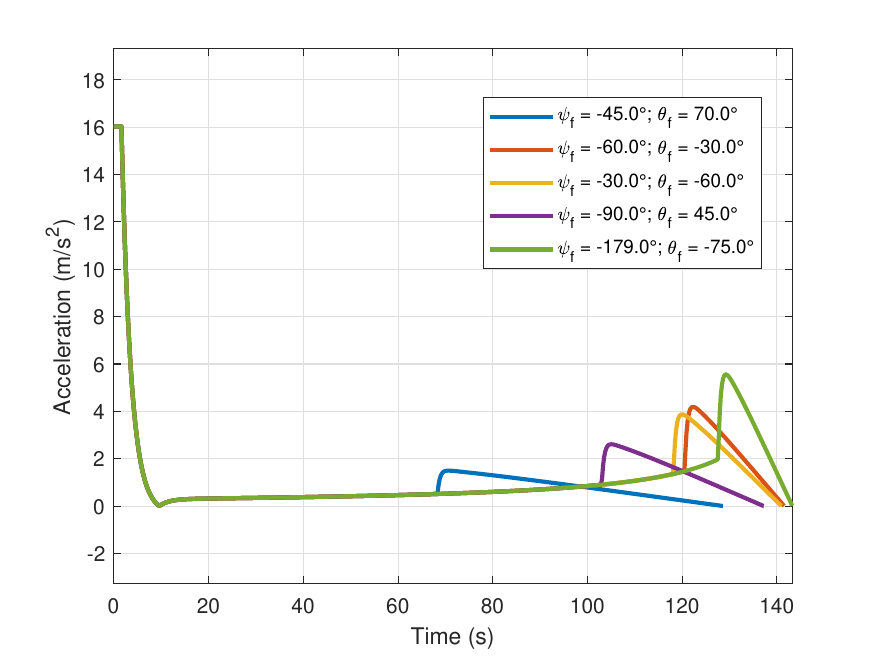}
    \label{s1:accn}
    }
    \subfloat[\centering Look Angle]{
    \includegraphics[width=0.5\linewidth]{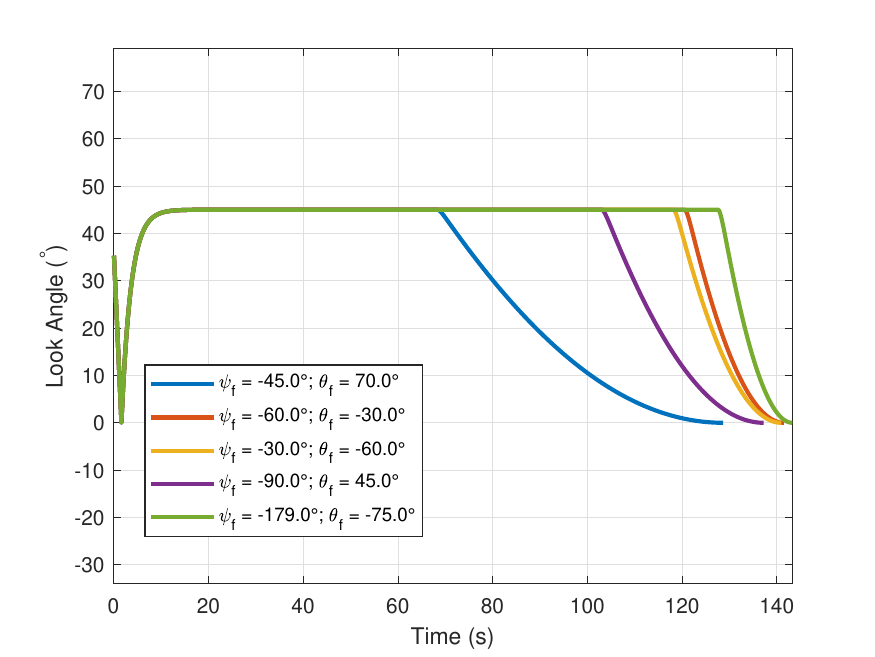}
    \label{s1:sgm}
    }\\
    \subfloat[\centering  Azimuth Angle]{
    \includegraphics[width=0.5\linewidth]{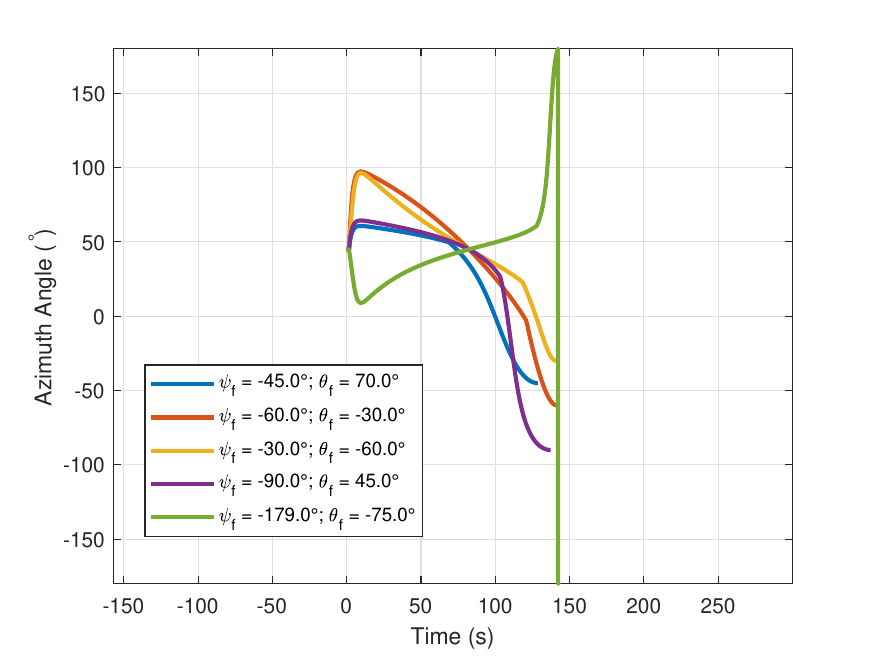}
    \label{s1:azi}
    }
    \subfloat[\centering Elevation Angle]{
    \includegraphics[width=0.5\linewidth]{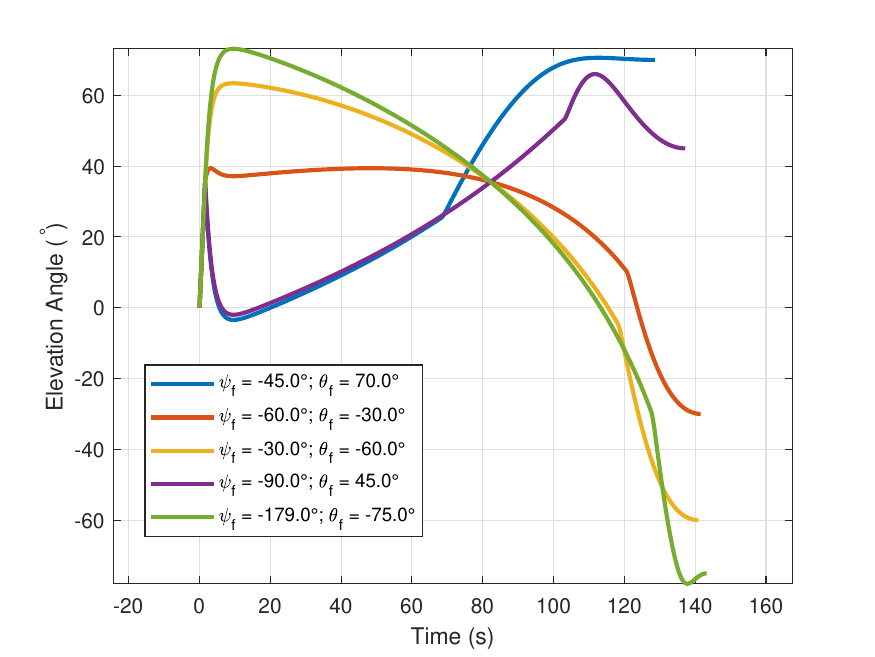}
    \label{s1:ele}
    }\\
    \caption{Simulation results of 3-Staged 3D IACG Law - Varying Impact Angles}
    \label{s1}
\end{figure}

As depicted in Fig.~\ref{s1:traj}, successful interception is achieved for all prescribed terminal angles. The lateral acceleration profiles in Fig.~\ref{s1:accn} remain perfectly continuous across all phase transitions, validating the virtual target placement strategy. The seeker look angle in  Fig.~\ref{s1:sgm} exhibits the designed three-stage behavior: monotonically decaying to zero in the primary plane, exponentially approaching $\sigma_{\text{max}}$ in the secondary plane, and finally converging to zero upon interception. Notably, the look angle remains strictly bounded by the $45^\circ$ limit throughout the engagement. Figs.~\ref{s1:azi} and \ref{s1:ele} show the azimuth and elevation angle histories corresponding to the different terminal impact angle cases.

\subsection{Results with Varying Launch Angles}

This scenario evaluates the robustness of the scheme against varying initial heading errors. The missile is launched from three distinct initial orientations: $(\psi_0, \theta_0)$ : $(20^\circ, 70^\circ)$, $(60^\circ, 30^\circ)$, and $(80^\circ, 10^\circ)$, with terminal impact angles $(\psi_f, \theta_f)$: $(-45^\circ, 35.26^\circ)$.

\begin{figure}[h!]
    \centering
    \subfloat[\centering Trajectory]{
    \includegraphics[width=0.6\linewidth]{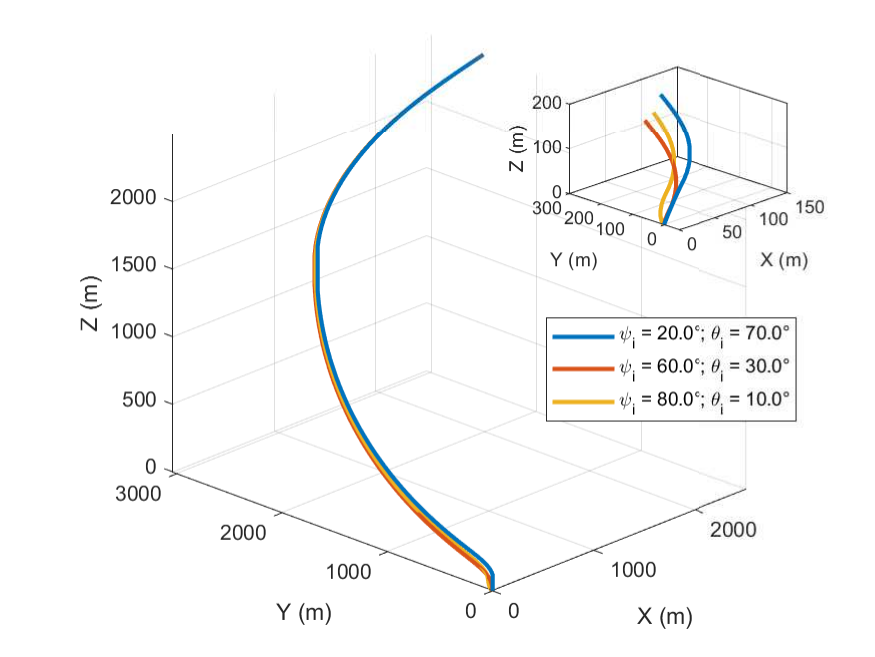}
    \label{s2:traj}
    }\\
    \subfloat[\centering Acceleration]{
    \includegraphics[width=0.5\linewidth]{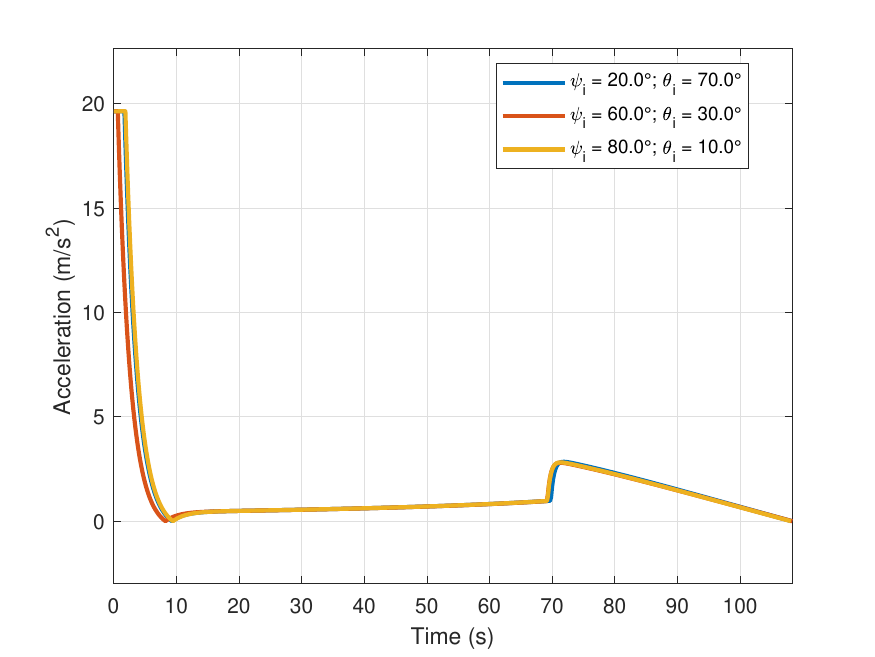}
    \label{s2:accn}
    }
    \subfloat[\centering Look Angle]{
    \includegraphics[width=0.5\linewidth]{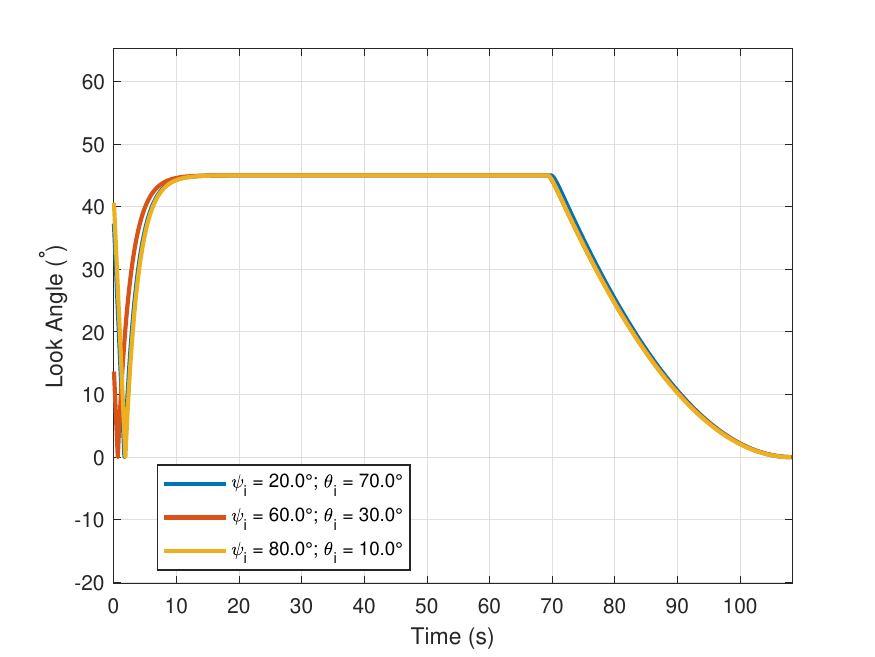}
    \label{s2:sgm}
    }\\
    \subfloat[\centering Azimuth Angle]{
    \includegraphics[width=0.5\linewidth]{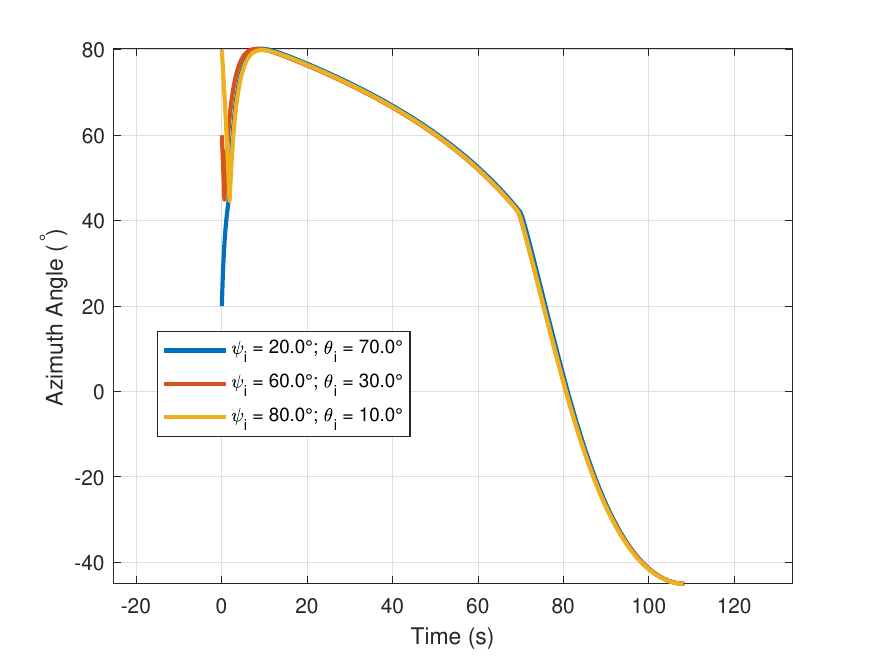}
    \label{s2:azi}
    }
    \subfloat[\centering Elevation Angle]{
    \includegraphics[width=0.5\linewidth]{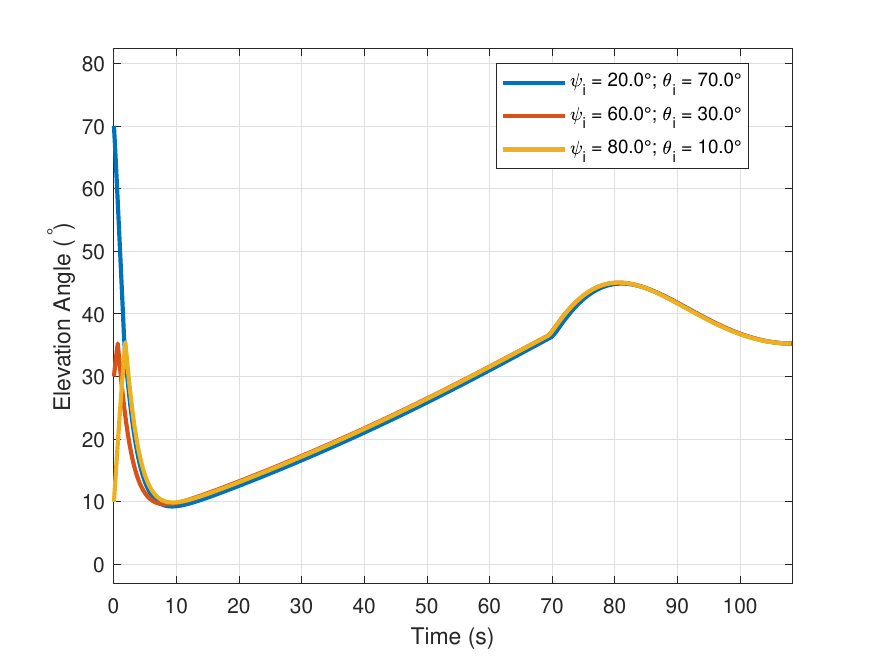}
    \label{s2:ele}
    }\\
    \caption{Simulation results of 3-Staged 3D IACG Law - Varying Launch Angles}
    \label{s2}
\end{figure}

Fig.~\ref{s2} confirms that the missile successfully intercepts the stationary target for all initial conditions. As the terminal constraints dominate the secondary planar engagement, the spatial trajectories in Fig.~\ref{s2:traj} nearly coincide after the initial alignment phase. Acceleration continuity is preserved, as shown in Fig.~\ref{s2:accn}, and the seeker FOV constraint is strictly satisfied, with the look angle remaining below $\sigma_{\text{max}}$, as illustrated in Fig.~\ref{s2:sgm}. Furthermore, Figs.~\ref{s2:azi} and~\ref{s2:ele} demonstrate that the desired terminal impact angles are consistently achieved regardless of the initial launch orientation.

\subsection{Comparison Study}
\begin{figure}[h!]
    \centering
    \subfloat[\centering Trajectory]{
    \includegraphics[width=0.6\linewidth]{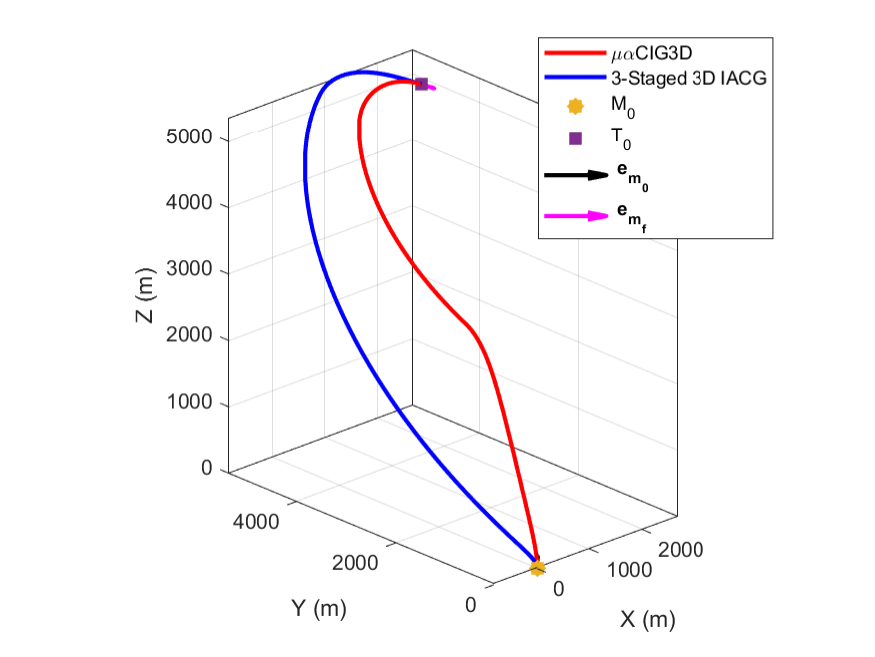}
    \label{s3:traj}
    }\\
    \subfloat[\centering Acceleration]{
    \includegraphics[width=0.5\linewidth]{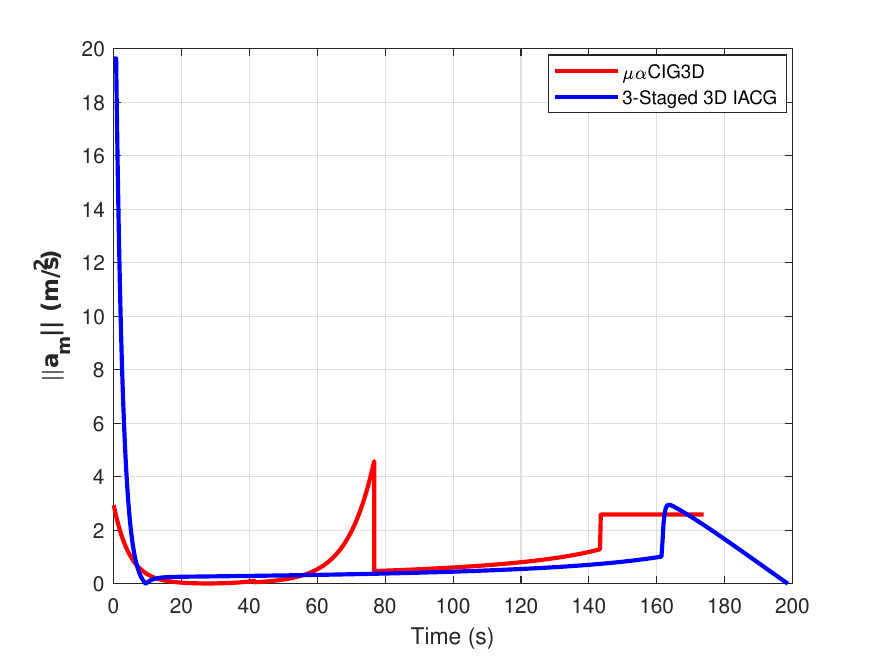}
    \label{s3:accn}
    }
    \subfloat[\centering Look Angle]{
    \includegraphics[width=0.5\linewidth]{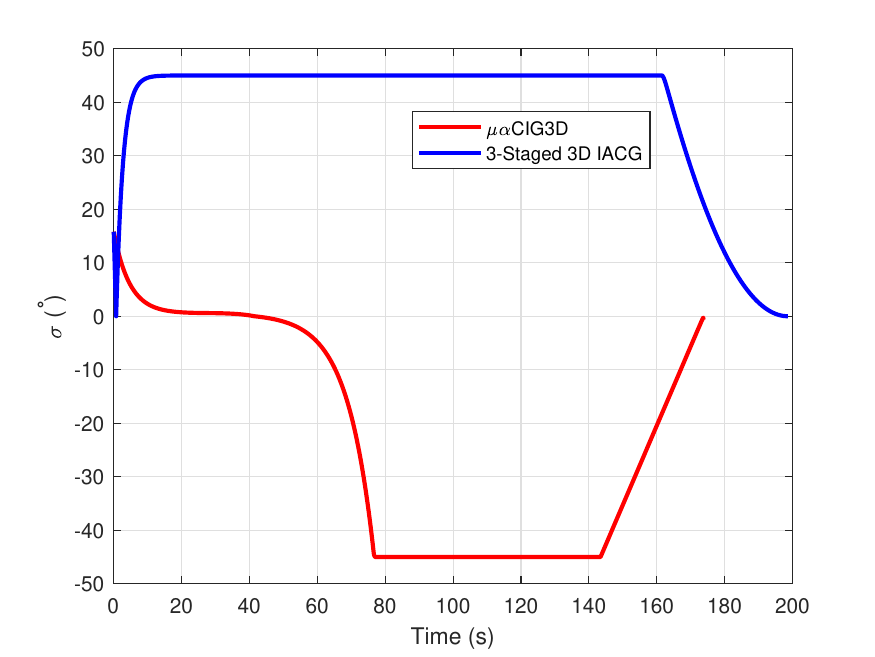}
    \label{s3:sgm}
    }\\
    \subfloat[\centering Azimuth Angle]{
    \includegraphics[width=0.5\linewidth]{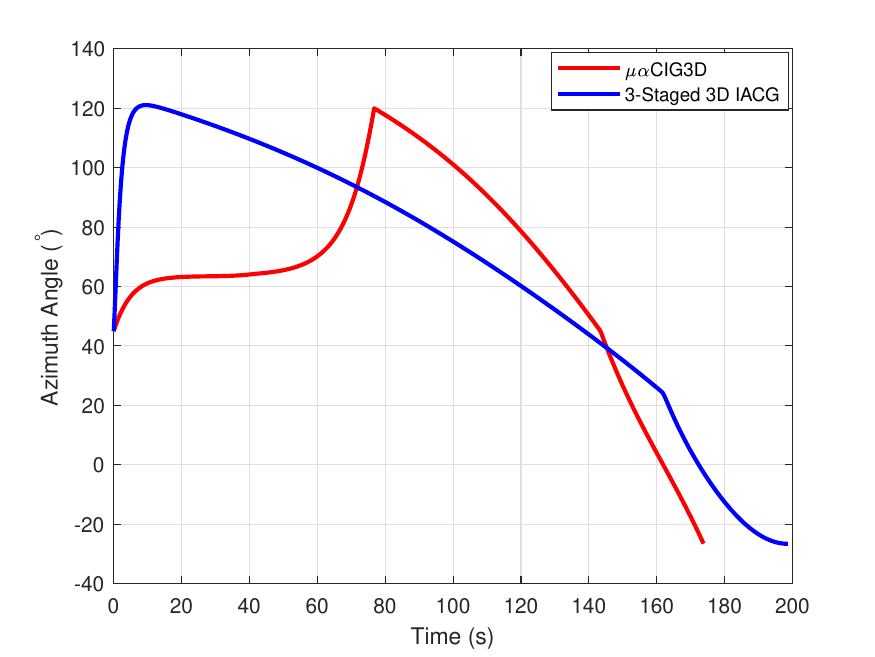}
    \label{s3:azi}
    }
    \subfloat[\centering Elevation Angle]{
    \includegraphics[width=0.5\linewidth]{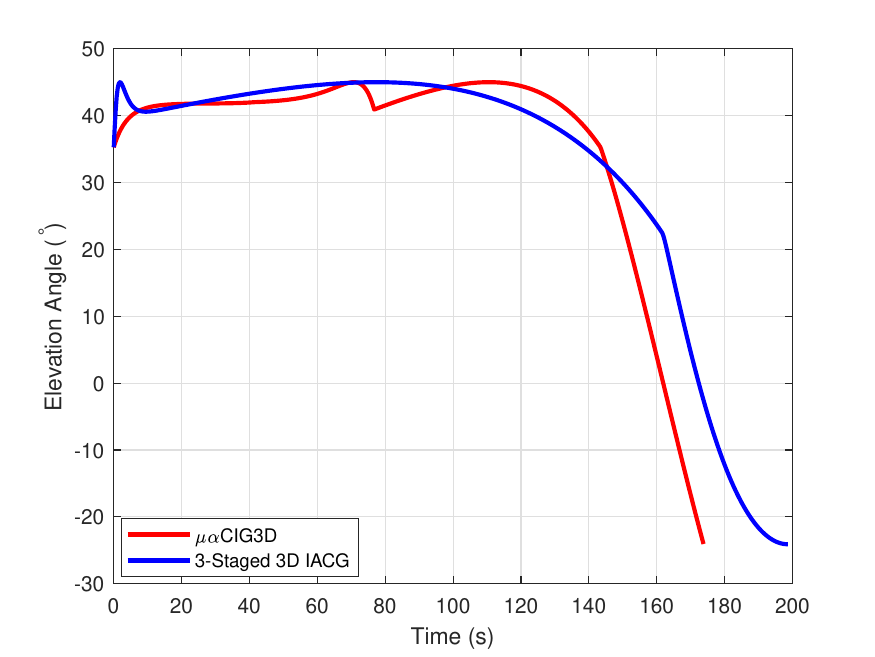}
    \label{s3:ele}
    }\\
    \caption{Comparative results}
    \label{s3}
\end{figure}

The proposed formulation is benchmarked against the composite PN-based $\mu\alpha\text{CIG3D}$ law \cite{ghosh2017unmanned}. For a representative engagement, the target is located at $[2500, 5000, 5000]^T$ m, and the missile is initialized with a velocity direction corresponding to $(\psi_0, \theta_0) = (45^\circ,35.26^\circ)$, with a commanded terminal impact angle of $(\psi_f, \theta_f) = (-45^\circ,-45^\circ)$. While both guidance laws achieve successful interception, as shown in Fig.~\ref{s3:traj}, the $\mu\alpha\text{CIG3D}$ formulation exhibits multiple severe discontinuities in the lateral acceleration command, as illustrated in Fig.~\ref{s3:accn}. In practical implementations, such abrupt variations would induce undesirable structural transients in the airframe. In contrast, the proposed 3D IACG law yields a smooth and continuous acceleration profile. Furthermore, Figs.~\ref{s3:azi} and~\ref{s3:ele} demonstrate that the desired terminal impact angles are consistently achieved by both guidance laws.

To quantify efficiency, the total control effort (evaluated as $\int a_M^2 \,dt$) is compared across six diverse terminal impact geometries. As summarized in Table~\ref{table1}, the proposed scheme consistently demands lower total control effort than $\mu\alpha\text{CIG3D}$ across all evaluated scenarios, highlighting the energetic efficiency of the continuous TVBPN transition over discrete composite formulations.

\begin{table}[h]
\caption{Comparison of Total Control Effort for Different Initial Angle Cases}
\label{table1}
\centering
\begin{tabular}{lccc}
\toprule
Method & $\psi$ (deg) & $\phi$ (deg) & Total control effort ($\mathrm{m^2/s^3}$) \\
\midrule
3-Staged 3D IACG  
& \multirow{2}{*}{-45} 
& \multirow{2}{*}{-45} 
& 1928.40 \\
$\mu\alpha$ CIG3D 
&  
&  
& 2054.20 \\
\midrule
3-Staged 3D IACG  
& \multirow{2}{*}{-60} 
& \multirow{2}{*}{-30} 
& 1954.08 \\
$\mu\alpha$ CIG3D 
&  
&  
& 2099.55 \\
\midrule
3-Staged 3D IACG  
& \multirow{2}{*}{-30} 
& \multirow{2}{*}{-60} 
& 1923.86 \\
$\mu\alpha$ CIG3D 
&  
&  
& 2044.83 \\
\midrule
3-Staged 3D IACG  
& \multirow{2}{*}{-90} 
& \multirow{2}{*}{-45} 
& 2173.96 \\
$\mu\alpha$ CIG3D 
&  
&  
& 2491.97 \\
\midrule
3-Staged 3D IACG  
& \multirow{2}{*}{-135} 
& \multirow{2}{*}{-60} 
& 2299.14 \\
$\mu\alpha$ CIG3D 
&  
&  
& 2704.93 \\
\midrule
3-Staged 3D IACG  
& \multirow{2}{*}{-180} 
& \multirow{2}{*}{-75} 
& 2089.15 \\
$\mu\alpha$ CIG3D 
&  
&  
& 2326.88 \\
\bottomrule
\end{tabular}
\end{table}

\subsection{Realistic Simulations}
To verify practical feasibility, the guidance scheme is applied to a realistic 6-DOF-equivalent point-mass missile model incorporating aerodynamic drag, gravity, and thrust profile variations. The governing equations for the time-varying velocity $V_M$, atmospheric density, and aerodynamic coefficients are detailed in the Appendix, adopted from \cite{kee1998near}. The missile is launched at $300$ m/s along $(\psi_0, \theta_0)$ : $(45^\circ, 0^\circ)$ toward a target at $[2500, 2500, 2500]^T$ m. Three distinct terminal impact angle cases are considered to evaluate the guidance performance, defined by $(\psi_f, \theta_f)$: $(-60^\circ, 30^\circ)$, $(-30^\circ,-60^\circ)$, and $(-179^\circ,-75^\circ)$.

\begin{figure}[h!]
    \centering
    \subfloat[\centering Trajectory]{
    \includegraphics[width=0.5\linewidth]{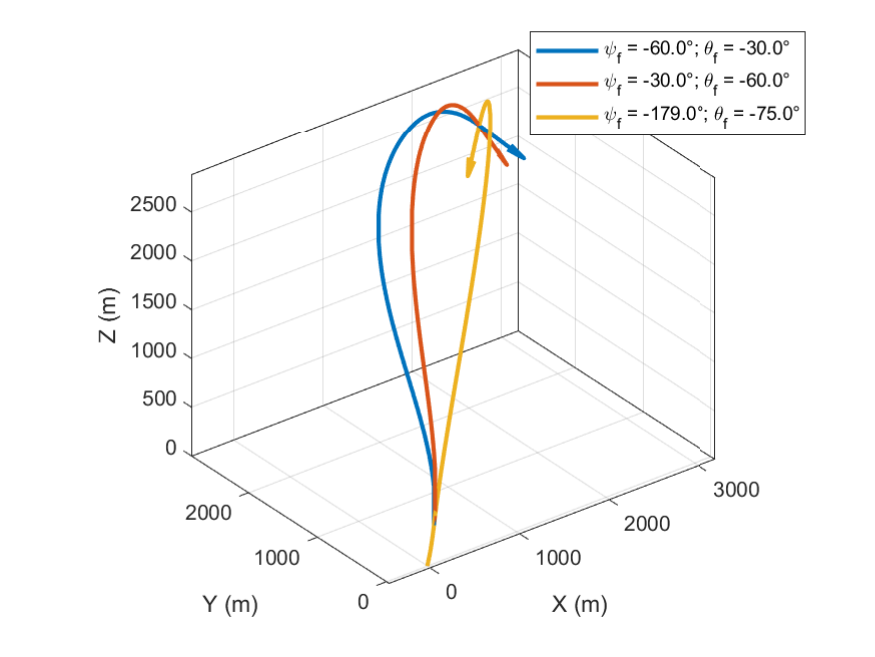}
    \label{s4:traj}
    }
    \subfloat[\centering Acceleration]{
    \includegraphics[width=0.5\linewidth]{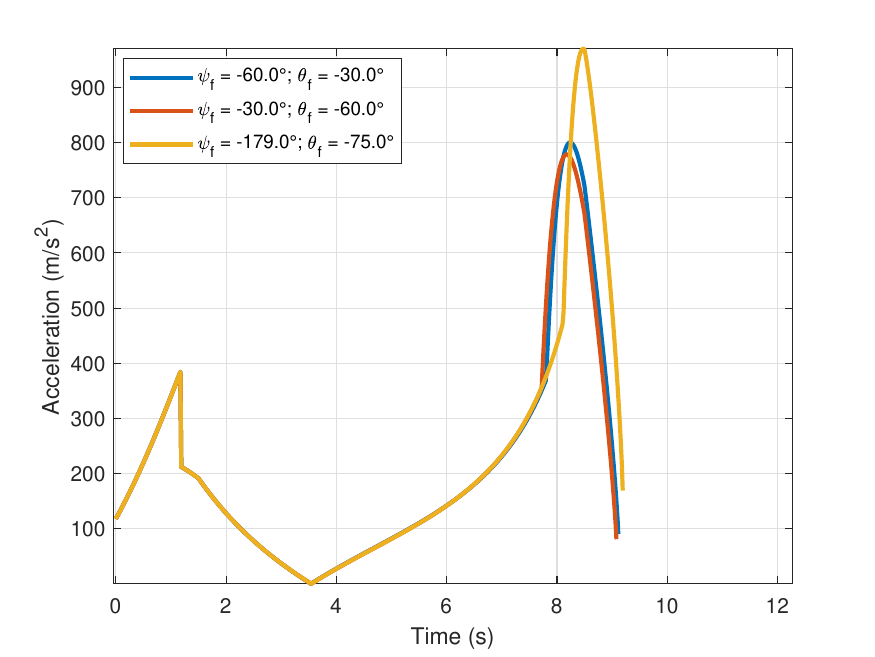}
    \label{s4:accn}
    }\\

    \subfloat[\centering Velocity]{
    \includegraphics[width=0.5\linewidth]{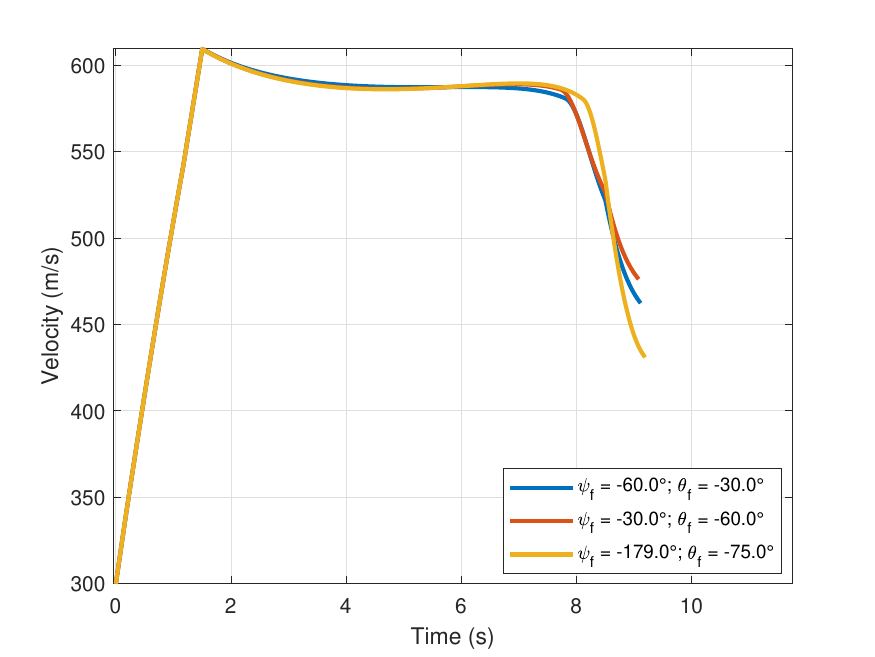}
    \label{s4:vel}
    }
    \subfloat[\centering Look Angle]{
    \includegraphics[width=0.5\linewidth]{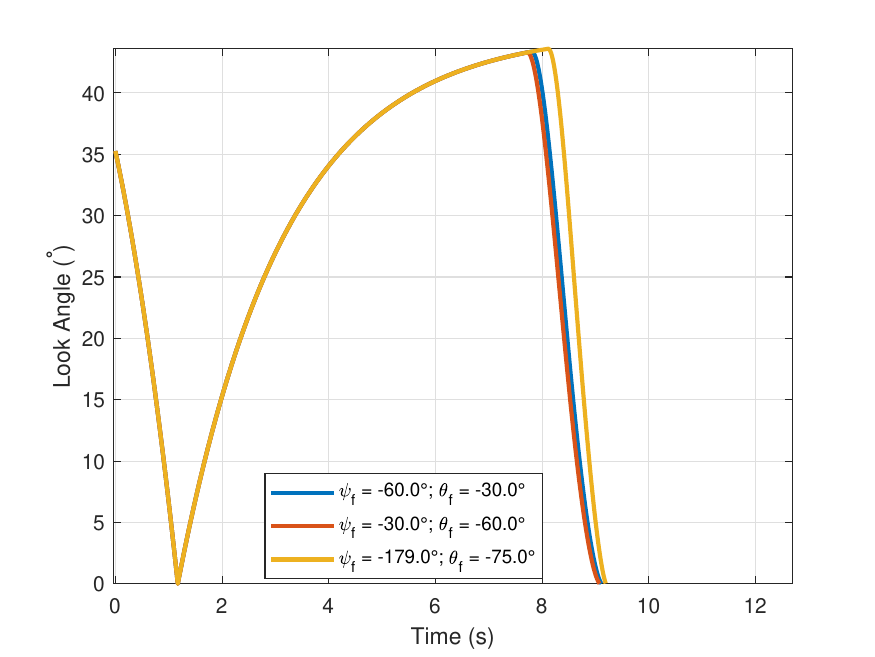}
    \label{s4:sgm}
    }\\

    \subfloat[\centering Azimuth Angle]{
    \includegraphics[width=0.5\linewidth]{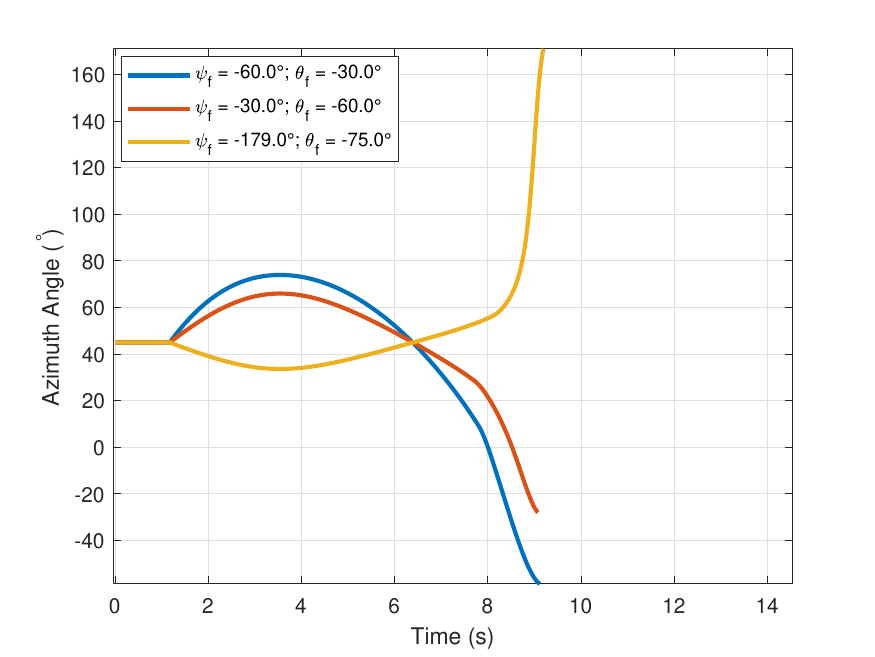}
    \label{s4:azi}
    }
    \subfloat[\centering Elevation Angle]{
    \includegraphics[width=0.5\linewidth]{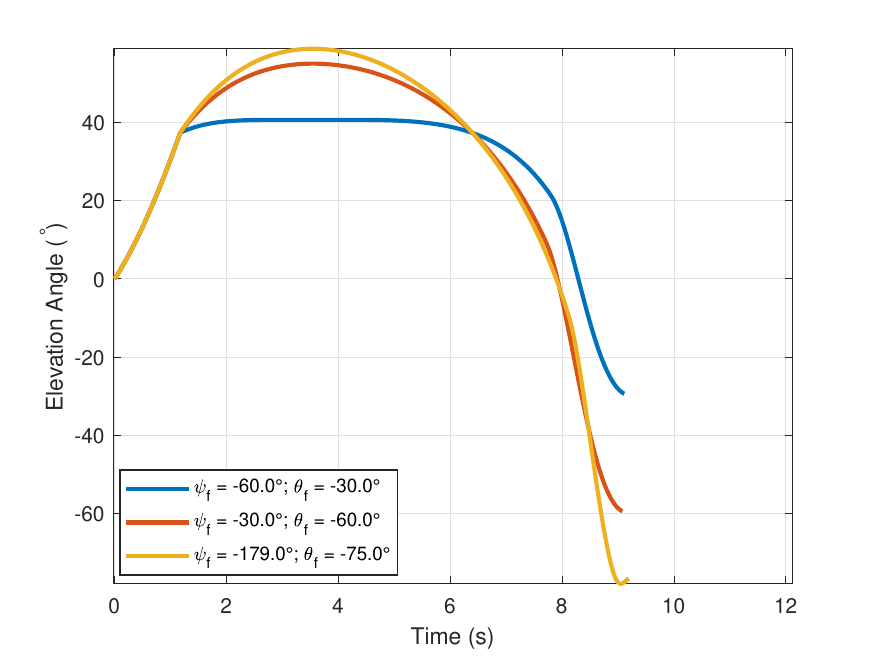}
    \label{s4:ele}
    }\\
    \caption{Realistic Simulation results}
    \label{s4}
\end{figure}

Fig.~\ref{s4:traj} shows the trajectories for three different cases, while the corresponding lateral acceleration profiles are presented in Fig.~\ref{s4:accn}. Despite the highly nonlinear velocity variation induced by aerodynamic drag and gravity, as shown in Fig.~\ref{s4:vel}, the guidance scheme successfully enforces the prescribed terminal impact angles, as illustrated in Figs.~\ref{s4:azi} and~\ref{s4:ele}. Crucially, the seeker look angle, shown in Fig.~\ref{s4:sgm}, remains bounded below the $45^\circ$ FOV limit. Although the lateral acceleration profiles in Fig.~\ref{s4:accn} exhibit minor deviations from the ideal constant profile during the primary phase due to time-varying speed, the transition behavior remains smooth, confirming the robustness of the proposed scheme under realistic engagement conditions.

\section{CONCLUSIONS}
\label{sec:concl}

A novel three-dimensional impact angle control guidance (IACG) law is developed to achieve specified terminal interception geometries while strictly enforcing seeker field-of-view (FOV) limits and guaranteeing lateral acceleration continuity. By decoupling the spatial engagement into two sequential planar phases, the scheme first employs a pure proportional navigation phase to align the velocity vector of the missile with the line-of-sight. The engagement then transitions to a secondary plane, where a modified two-stage time-varying biased proportional navigation (TVBPN) law actively shapes the terminal trajectory to satisfy the prescribed impact angle without violating FOV constraints. To eliminate the acceleration discontinuities typically associated with multi-stage guidance, an analytically positioned virtual target is introduced during the primary phase. Extensive nonlinear simulations, including comparative analysis and realistic missile dynamics, demonstrate the energetic efficiency, robustness, and practical feasibility of the proposed formulation.
 
\renewcommand{\theequation}{A-\arabic{equation}}
\setcounter{equation}{0}  

\section*{APPENDIX}
\subsection{Realistic Vehicle Model Parameters}

The 6-DOF-equivalent point-mass kinematics and aerodynamic properties utilized for the realistic simulations in Section~\ref{sec:sims}.D are governed by the following equations, adopted from \cite{kee1998near}. The time-varying missile velocity, $V_M$, is determined by the thrust $T$, aerodynamic drag $D$, mass $m$, and elevation angle $\theta$:
\begin{equation}\label{A1}\dot{V}_M = \frac{T - D}{m} - g \sin\theta
\end{equation}
where, $g$ is the gravitational acceleration. The aerodynamic drag $D$ and required lift coefficient $C_L$ to sustain the commanded lateral acceleration $a_M$ are formulated as:
\begin{equation}
\label{A2}
D = Q C_D S, \quad C_L = \frac{m a_M}{Q S}
\end{equation}
where, $S=1 \text{ m}^2$ is the reference area, $C_D$ is the total drag coefficient, and the dynamic pressure $Q$ is given by $Q = \frac{1}{2} \rho V_M^2$. The total drag coefficient is a function of the zero-lift drag coefficient $C{D_0}$ and the induced drag coefficient $K$:\begin{equation}\label{A3}C_D = C_{D_0} + K C_L^2\end{equation}
The atmospheric temperature $\textup{T}$ (in Kelvin), speed of sound $a$, Mach number $M$, and atmospheric density $\rho$ (in $\text{kg/m}^3$) are modeled as functions of the altitude $h$ (in meters):
\begin{equation}
\label{A4}
\textup{T} = 
\begin{cases} 
(288.16 - 0.0065h)\textup{ K}& h < 11000 \textup{ m} \\
216.66\textup{ K} & h \ge 11000
\end{cases}
\end{equation}

\begin{equation}
\label{A5}
a = \sqrt{\gamma R \textup{T}}, \quad M = \frac{V_M}{a}
\end{equation}

\begin{equation}
\label{A6}
\rho(h) = 1.15579 - 1.058 \times 10^{-4}h + 3.725 \times 10^{-9}h^2 - 6.0 \times 10^{-14}h^3
\end{equation}
where, $\gamma=1.4$ is the heat capacity ratio and $R=287.05 \text{ J/(kg}\cdot\text{K)}$ is the universal gas constant. The aerodynamic coefficients $C_{D_0}$ and $K$ vary piecewise with the Mach number: 
\begin{equation}
\label{A7}
C_{D_0} = 
\begin{cases} 
0.02 & M < 0.93 \\
0.02 + 0.2(M - 0.93) & M < 1.03 \\
0.04 + 0.06(M - 1.03) & M < 1.10 \\
0.0442 - 0.007(M - 1.10) & M \ge 1.10
\end{cases}
\end{equation}

\begin{equation}
\label{A8}
K = 
\begin{cases} 
0.2 & M < 1.15 \\
0.2 + 0.246(M - 1.15) & M \ge 1.15
\end{cases}
\end{equation}
Finally, the  thrust profile $T$ (in Newtons) of the vehicle and mass $m$ (in kg) are modeled as functions of the flight time $t$ (in seconds):
\begin{equation}
\label{A9}
T = 
\begin{cases} 
33000 & 0 \le t \le 1.5 \\
7500 & 1.5 \le t \le 8.5 \\
0 & 8.5 \le t
\end{cases}
\end{equation}

\begin{equation}
\label{A10}
m = 
\begin{cases} 
(135 - 14.53t) & 0 \le t \le 1.5 \\
(113.205 - 3.31t) & 1.5 \le t \le 8.5 \\
90.035 & 8.5 \le t
\end{cases}
\end{equation}
where, $t$ is the time from the launch.

\end{document}